\documentstyle[12pt]{article}
\newcommand{\beq}{\begin{equation}}
\newcommand{\eeq}{\end{equation}}
\newcommand{\beqa}{\begin{eqnarray}}
\newcommand{\eeqa}{\end{eqnarray}}
\newcommand{\ba}{\begin{array}}
\newcommand{\ea}{\end{array}}

\newcommand{\A}{\alpha}

\newcommand{\D}{\delta}

\newcommand{\La}{\Lambda}

\newcommand{\lm}{\lambda}
\newcommand{\s}{\sigma}

\renewcommand{\thefootnote}{\fnsymbol{footnote}}

\begin{document}
\begin{titlepage}
\begin{flushright}
{\tt hep-th/9903260} \\
March, 1999
\end{flushright}
\vspace{0.5cm}
\begin{center}
{\Large \bf 
A Note on Higher Dimensional Instantons \\
and Supersymmetric Cycles\footnote{
Talk presented at the workshop on
\lq\lq Gauge Theory and
Integrable Models\rq\rq\ 
(YITP, Kyoto),  January 26-29, 1999.}
} %
\lineskip .8em
\vskip2.5cm
{\large Hiroaki Kanno}
\vskip 1.5em
{\large\it Department of Mathematics, Hiroshima University \\
Higashi-Hiroshima, 739-8526, Japan}
\end{center}
\vskip2cm
\begin{abstract}
We discuss instantons in dimensions higher than four. 
A generalized self-dual or anti-self-dual
instanton equation in $n$-dimensions can be defined in terms of 
a closed $(n-4)$ form $\Omega$ and it was recently
employed as a topological gauge fixing condition in higher
dimensional generalizations of cohomological Yang-Mills theory. 
When $\Omega$ is a calibration
which is naturally introduced on the manifold of special holomony,
we argue that higher dimensional instanton may be 
locally characterized as a family of
four dimensional instantons over a supersymmetric $(n-4)$ cycle
$\Sigma$ with respect to the calibration $\Omega$.
This is an instanton configuration on the total space of the
normal bundle $N(\Sigma)$ of the submanifold $\Sigma$ 
and regarded as a natural generalization of point-like instanton
in four dimensions that plays a distinguished role in a compactification
of instanton moduli space.

\end{abstract}
\end{titlepage}
\baselineskip=0.55cm

\renewcommand{\thefootnote}{\arabic{footnote}}
\setcounter{footnote}{0}


\section{Introduction}

Instantons or (anti-)self-dual connections in four dimensions
are interesting and important objects both in mathematics
and physics. In this article we discuss higher dimensional
generalization. One of motivations to consider higher 
dimensional instantons comes from recent developments
in string dualities and $M$-theory, where we obtain
low energy effective (supersymmetric) gauge 
theories in diverse dimensions. We are also motivated by
its use in topological (or more precisely cohomological) 
gauge theories in higher dimensions. We can employ higher
dimensional instanton equation as a topological gauge
fixing condition in BRST quantization 
of topological action in eight and other dimensions
\cite{BKS1},\cite{BKS2},\cite{AOS}. Consequently,
we expect such cohomological gauge theory explores the
moduli space of higher dimensional instantons.
Supersymmetry plays a key role in all of
these subjects.

Our higher dimensional instanton equation is
\beq
\Omega \wedge F \pm (* F) =0~, \label{inst}
\eeq
where $\Omega$ is a closed $(n-4)$ form on an $n$ dimensional 
Riemannian manifold. (See the next section for details.)
We will call a solution either anti-self-dual or self-dual
according to the choice of the sign in (\ref{inst}).
We argue that, when $\Omega$ defines a calibration,
higher dimensional instantons are locally characterized
as a family of four dimensional instantons over an $(n-4)$ dimensional
cycle $\Sigma$. The generalized instanton equation (\ref{inst}) 
requires that the cycle $\Sigma$ should be supersymmetric. 
(The idea of supersymmetric (SUSY) cycles and the relation to
the calibration $\Omega$ are reviewed in section 3.)
More precisely, by rescaling the transverse coordinates to $\Sigma$,
we will see that higher dimensional instantons can be reduced
to those on the total space of the normal bundle $N(\Sigma)$ 
of the SUSY cycle $\Sigma$. Note that the fibre of $N(\Sigma)$ is
the four dimensional Euclidean space ${\bf R}^4$.
Such a scaled instanton is the most natural generalization of point-like
(or ideal) instanton in four dimensions where
the SUSY cycle is just a set of points. We expect that
it will play a crucial role in an attempt at compactifying
the moduli space of higher dimensional instantons
\cite{Nak},\cite{DT},\cite{Tian}.

 We already know a trivial example of a family of 
instantons. It is the gauge five-brane solution in the heterotic
string theory, where four dimensional Yang-Mills instanton
lives in the transverse direction to the world-volume of 
flat five-brane \cite{Sto}. 
This is a four dimensional instanton trivially embedded to
ten dimensional space-time and should not be regarded 
as a genuine higher dimensional instanton.
Our proposal on higher dimensional instantons modifies it in two aspects. 
Firstly, our supersymmetric cycle $\Sigma$ is curved in general and,
especially, can be compact, while the world-volume of
the gauge five-brane solution is flat and non-compact.
Secondly and more significantly, the moduli (or the size and shape) of
instanton in general depends on the coordinates of
the base space $\Sigma$ and hence our solution is not a direct product of 
four dimensional instanton and the world volume of the brane.

The connection of cohomological Yang-Mills theories 
based on the generalized instanton equations
to the D-branes and the calibrated geometry has
been discussed in e.g. \cite{AFOS},\cite{BT}. 
In these literatures the entire
manifold (e.g. a manifold of $Spin(7)$ holonomy
for the octonionic instanton) was identified
with the world volume of D-brane or a supersymmetric
cycle in ten or eleven dimensional space-time.
This is very natural viewpoint to see
the fact that the cohomological Yang-Mills theory in eight
dimensions is in fact a dimensional
reduction of ten-dimensional super Yang-Mills theory.
However, we emphasize that our viewpoint on supersymmetric cycles 
is rather different from the above story. In this paper
the supersymmetric cycle is not the entire manifold,
but a calibrated submanifold in the manifold
of special holonomy on which the gauge theory is defined.
The idea of branes within branes might reconcile these two
pictures \cite{Dou}.


\section{Generalized Self-Dual Instanton Equation}

The instanton equation in four dimensions is 
the (anti-)self-dual condition on the curvature two-form $F$,
which is quite intrinsic to four dimensions.
We can generalize the (anti-)self-duality to higher dimensions
in the following way. Let $(M,g)$ be an $n$ dimensional
manifold with a Riemannian metric $g$. We assume there
exists a closed $(n-4)$ form $\Omega$. Then the generalization
in the self-dual case is
\beq
\Omega \wedge F =  (* F)~, \label{OASD}
\eeq
where $*$ is the Hodge dual operator defined by the metric.
In \cite{CDFN} this equation was first appeared in the form
\beq
F^{\mu\nu}~=~\frac{1}{2} T^{\mu\nu\rho\sigma} F_{\rho\sigma}~,
\eeq
with a totally anti-symmetric tensor $T^{\mu\nu\rho\sigma}$.
If we think of $T^{\mu\nu\rho\sigma}$ as components
of a four-form $T$, then the relation of these conditions 
is simply given by $T = (*~\Omega)$. We take the equation (\ref{OASD}) as our
definition of self-duality, 
since $\Omega$ is geometrically more natural object
as we will see in the following.
A solution to the generalized self-dual equation is necessarily 
a solution to the Yang-Mills equation,
because
\beq
D_A * F = D_A (\Omega \wedge F) = \pm \Omega \wedge D_A F = 0~,
\eeq
where we have used $d\Omega = 0$ and the Bianchi identity.
For a configuration that satisfies the generalized self-dual condition,
the total action is given by a topological density;
\beq
\int {\hbox {Tr}} (* F \wedge F) = 
 \int \Omega \wedge {\hbox {Tr}} (F \wedge F)~.
\eeq

A typical example of the generalized instanton equation is
the equation of Donaldson-Uhlenbeck-Yau (DUY)
on a K\"ahler manifold in six dimensions, 
where $\Omega$ is the K\"ahler two form $\omega$. 
To reproduce the DUY equation, it is appropriate
to take the anti-self-dual condition.
Since $\omega^3$ gives the volume form up to the normalization, we have
$ * \omega = \omega \wedge \omega$.
The generalized anti-self-dual equation implies
\beq
\omega \wedge (*F) =  - (\omega \wedge \omega) \wedge F 
= - (*\omega) \wedge F~.
\eeq
Hence, we have the DUY equation;
\beq
\omega \wedge (*F) = 0~,  \quad {\hbox {or}}
\quad  \omega_{\mu\nu} F^{\mu\nu} = 0~.
\eeq
We have other examples of the $G_2$ instanton equation
in seven dimensions where $\Omega$ is the associative
three form and also the octonionic instanton equation in
eight dimensions where $\Omega$ is the Cayley 
four form $\Phi$.

When one first tries to construct an example of higher dimensional
instantons on the flat Euclidean space, he will soon encounter the
following problem.
Namely it has been argued that there are no finite action solutions
to the Yang-Mills equation in dimensions other than four.
One of the reasonings goes as follows\footnote{
This type of argument is known as the Derrick's theorem 
\cite{Der},\cite{Col},\cite{Wei}.};
The energy-momentum tensor of the Yang-Mills theory
\beq
T_{\mu\nu}~=~{\hbox {Tr}} \left(
F_{\mu\lm} F_\nu^{\ \lm} - \frac{1}{4} g_{\mu\nu}
F_{\lm\rho} F^{\lm\rho} \right)~,
\eeq
is conserved due to the equation of motion and the Bianchi
identity. Hence we have
\beq
\int d^n x~T_{\mu\nu}(x) =0~,
\eeq
where no surface term appears, if the action is supposed to be finite.
Taking trace, we obtain
\beq
(4-n) \int d^nx~{\hbox {Tr}} \left(
F_{\mu\nu} F^{\mu\nu} \right) =0~.
\eeq
Since the integral is nothing but the standard Yang-Mills action
itself, we come to the conclusion. This argument shows that
the problem is closely related to the fact that
the standard Yang-Mills action is conformal invariant
only in four dimensions. We have to accept that we can
find no self-dual instanton with finite action on the flat
Euclidean space ${\bf R}^n$ or its conformal compactification,
the $n$-dimensional sphere $S^n$. 
Thus there are no higher dimensional analogues
of the basic (BPST) instanton on $S^4$.
Point-like (ideal) instantons are important objects
in the compactification of the instanton moduli space
in four dimensions \cite{DK}. They are modeled after 
a superposition of BPST instantons. Therefore, to consider
a possible compactification of the moduli space of 
higher dimensional instantons,
it is desirable to have
a substitute in higher dimensions 
for the BPST instanton in four dimensions.
To look for such an object we will try to make a local splitting
of the total space into a four dimensional part and the remaining
$(n-4)$ dimensional one. This splitting is motivated by the conformal invariance
of the four dimensional Yang-Mills action. 
Point-like instantons can be obtained
by a conformal rescaling in four dimensions. Taking this fact into account,
we make use of the rescaling in the four dimensional part in 
the splitting. In section 4 we will work out 
a possible configuration of instanton that is stable under this scaling
and find the four dimensional self-dual instantons together 
with an $(n-4)$ dimensional supersymmetric cycle.

To elucidate the above idea of splitting, let us look at
the case of five dimensions, where $\Omega$ is a one form
\beq
\Omega = {\bf n} = n_\mu dx^\mu~.
\eeq
Since
\beq
{\bf n} \wedge (*F) = {\bf n} \wedge {\bf n} \wedge F = 0~,
\eeq
we see that the curvature $F$ is transverse to ${\bf n}$ in the sense that
\beq
n_\mu F^{\mu\nu}=0
\eeq
Futhermore, the generalized self-dual instanton equation reduces to
four dimensional instanton equation on the transverse 
subspace to the vector $n_\mu$. Thus in five dimensions $\Omega$
defines the normal direction to the four dimensional plane where
an instanton \lq\lq lives\rq\rq. In general we have to make it clear
what is the geometrical meaning of $\Omega$.
In the above argument the normal direction is defined at each point. 
On a curved manifold we would have a curve whose cotangent
vector is given by the one form $\Omega = n_\mu dx^\mu$.
We may think of this curve as a world line of a superparticle.
Thus in higher dimensions we are naturally led to the idea of the
world volume of a super $p$-brane and supersymmetric cycles.

Before concluding this section we would like to point out an additional issue
in the energy-momentum tensor of higher dimensional self-dual
instantons. In four dimensions both self-duality and
anti-self-duality of the curvature imply the vanishing of
the energy-momentum tensor. This property allows us
to consider instantons on Ricci flat manifolds
such as $K3$ surface and ALE spaces without disturbing
Ricci-flatness. They are consistent with the Einstein equation.
Unfortunately, this nice feature is lost
in higher dimensions. If the energy momentum vanishes,
we have
\beq
g_{\mu\nu} T^{\mu\nu}~=~\frac{1}{4} (4-n) {\hbox {Tr}} \left(
F_{\mu\nu} F^{\mu\nu} \right) =0~.
\eeq
If $n \neq 4$, this means the action density must vanish and hence
it gives a trivial solution. We should examine a physical consistency
of higher dimensional instanton on a Ricci-flat manifold.


\section{Supersymmetric Cycles}

In this section we review the idea of supersymmetric cycles
following \cite{BBS} and \cite{BSV}.
Consider a world volume $\Sigma$ of $p$-brane with local
coordinates $(\sigma^a)$ and an embedding of $\Sigma$
into the superspace ${\cal SM}$ with bosonic and fermionic
coordinates $(X^\mu, \theta^\A)$. When $p=1$, this is 
a standard framework of the Green-Schwarz formulation of
superstring theory. We ask what is a supersymmetric configuration
that is purely bosonic. A supersymmetry transformation of $X(\s)$ is
automatically vanishing, since we have no fermionic backgrounds.
The non-trivial condition arises from the requirement of
the vanishing of SUSY variation of fermionic coordinates;
\beq
\D_{SUSY} \theta = \epsilon~,
\eeq
where $\epsilon$ is a global SUSY parameter of the space-time ${\cal SM}$.
As was pointed out in \cite{BBS} we have to take into accout
the kappa-symmetry of super $p$-brane action;
\beq
\D_\kappa \theta = P_+ \kappa(\s)~.
\eeq
The projection operator $P_\pm$ is defined by
\beq
P_\pm = \frac{1}{2} \left( 1 \pm
\frac{1}{(p+1) ! \sqrt{h}} \epsilon^{a_1 \cdots a_{p+1}}
\Pi_{a_1}^{\mu_1} \cdots \Pi_{a_{p+1}}^{\mu_{p+1}} 
\Gamma_{\mu_1 \cdots \mu_{p+1}} \right)~,
\eeq
where 
\beq
\Pi_a^\mu = \partial_a X^\mu - i \left( \bar\theta \Gamma^\mu
\partial_a \theta - \partial_a \bar\theta \Gamma^\mu \theta \right)
\label{vielbein}
\eeq
is a supercovariant vielbein and $\Gamma_{\mu_1 \cdots \mu_{p+1}}$ is
the totally anti-symmetric product of space-time gamma
matrices $\Gamma_\mu$. The induced metric on the world
volume is $h_{ab} = \Pi_a^\mu \Pi_b^\nu g_{\mu\nu}$ and
$h$ is its determinant. Since we are interested in purely
bosonic background, we can neglect the fermion bilinear
terms in (\ref{vielbein}). 

Now a $(p+1)$ cycle $\Sigma$ is supersymmetric, if it is invariant
under supersymmetry transformation modulo $\kappa$-symmetry.
For purely bosonic configurations this means that the translation
in fermionic direction has to be compensated by $\kappa$-symmetry;
\beq
^\exists\!\epsilon \in {\hbox{Im}}~P_+ \iff 
^\exists\!\epsilon \in {\hbox{Ker}}~P_-~. 
\eeq
We have
\beq
P_- \epsilon = 0~,
\eeq
for a SUSY cycle $\Sigma$.
Once the homology class of $(p+1)$ cycle is fixed, 
the supersymmetric cycles are volume minimizing.
We can see this property as follows;
for any $(p+1)$ cycle $\Sigma$, or a world volume of $p$-brane, we have
\beq
\int_\Sigma d^{p+1} \sigma \sqrt{h}~\overline{(P_- \epsilon)}
(P_- \epsilon) =
\int_\Sigma d^{p+1} \sigma \sqrt{h}~\overline{\epsilon}
(P_- \epsilon) \geq 0~,
\eeq
where the equality holds only for SUSY cycles.
Substituting the form of the projection $P_-$,
we obtain a lower bound
\beq
{\hbox{vol.}} (\Sigma) 
 = \int_\Sigma d^{p+1} \sigma \sqrt{h}~\overline{\epsilon}
\epsilon  \geq  \int_\Sigma X^* (\Omega)
\label{bound}
\eeq
where we have used a normalization
$\overline{\epsilon}\epsilon = 1$ and
\beq
\Omega = \left( \overline{\epsilon}~\Gamma_{\mu_1 \cdots \mu_{p+1}}
\epsilon \right) dX^{\mu_1} \wedge \cdots \wedge dX^{\mu_{p+1}}~,
\label{Omega}
\eeq
is a $(p+1)$ form on the space-time.
The right hand side of eq.(\ref{bound}) depends
only on the homology class of $\Sigma$ and
it gives the lower bound of the volume of 
the $(p+1)$ cycles with a fixed homology class.
In supersymmetric field theory a bound
of this type is called BPS bound. We see the
SUSY cycles are BPS saturated states.
For SUSY cycles that saturate the bound 
the volume form ${vol}_\Sigma$ is given by the pull-back
$X^*(\Omega)$ of the space-time $(p+1)$ form
$\Omega$ and this means
\beq
\Omega |_\Sigma = {vol}_\Sigma~, \label{calib}
\eeq
if we identify the $(p+1)$ cycle with its image in the space-time.

According to \cite{HL},\cite{Har},\cite{Mc}, 
a closed form that satisfies the inequality of the type (\ref{bound})
defines a calibration on the manifold. Then,
a cycle with the property (\ref{calib}) is called $\Omega$-calibrated
submanifold . 
The geometry of SUSY cycles is the theory of calibrated submanifolds in
differential
geometry. The space-time $(p+1)$ form $\Omega$ defined by
eq.(\ref{Omega}) is closed, if $\epsilon$ is covariantly constant.
Thus we are led to consider the
manifolds of special holonomy such as Calabi-Yau,
hyperK\"ahler, $G_2$ and $Spin(7)$ (or Joyce) manifolds
that allow a covariantly constant spinor.
It is known that all manifolds in this class are Ricci-flat.
In the following we assume $\epsilon$ is a covariantly
constant spinor, so that we may use the $(p+1)$ form $\Omega$ 
to define the generalized self-dual equation (\ref{OASD}).


\section{Topology of Higher Dimensional
Instantons}

Let us consider the generalized self-dual equation (\ref{OASD}) with a
calibration
$\Omega$ introduced in the last section.
The $(n-4)$ form $\Omega$ is used to define SUSY
cycles on an $n$-dimensional manifold $M^n$ of special holonomy.
Following the argument in section 2,
we will make a splitting of local coodinates
around a point $x \in M^n$ into \lq\lq transverse\rq\rq\ 
four dimensional ones $(y^1, \cdots, y^4)$ and the remaining
coordinates $(z^1, \cdots, z^{n-4})$. The transverse direction
may be regarded as a fibre in the normal bundle $N(\Sigma)$ 
over the $(n-4)$ dimensional base space $\Sigma$ 
with the coordinates $(z^1, \cdots, z^{n-4})$. 
At this stage we do not assume that the cycle $\Sigma$ is
supersymmetric. It will be derived from the generalized 
self-dual condition (\ref{OASD}).
We then consider a conformal rescaling along the transverse directions. 
This rescaling may be equivalent to a rescaling of the metric, which is
a standard trick in topological dimensional reduction \cite{BJSV},\cite{FKS}. 
But here we will not assume that the manifold $M^n$ is
a direct product, which is a usual ansatz in the procedure of dimensional
reduction.
Though our consideration will be restricted 
to a neighborhood of the point $x$ we chose, 
as we will discuss later our reduction is definitely
different from the dimensional reduction.
This partial rescaling of four dimensional fibre of $N(\Sigma)$ 
is made possible by the conformal
invariance of the Yang-Mills action in four dimensions.
In four dimensional case such a rescaling produces
a point-like instanton. Therefore, it is expected that the partial
rescaling enables us to see an analogue of point-like
instanton that is localized on the $(n-4)$ dimensional cycle $\Sigma$.
Under the rescaling of the transverse coordinates
\beq
y^i \longrightarrow \lm y^i~,
\eeq
the scaling of the curvature is
\beq
F = \lm^{-2} F_{yy} + \lm^{-1} F_{yz} + F_{zz}~,
\eeq
and the Yang-Mills action is transformed into
\beq
\int |F|^2 dydz = \int |F_{yy}|^2 dydz + \lm^2 \int |F_{yz}|^2 dydz
+ \lm^4 \int |F_{zz}|^2 dydz~.
\eeq
Thus to keep the action finite in the limit $\lm \rightarrow \infty$,
both the components $F_{yz}$ and $F_{zz}$ should be suppressed and
we have a reduced self-dual condition
\beq
\Omega \wedge F_{yy} =  * F_{yy}~, \label{transASD}
\eeq
in this limit. We can think of the equation (\ref{transASD}) as 
the condition satisfied by the higher dimensional analogue of point-like
instanton that is stable under the partial rescaling. 
To make the meaning of eq.(\ref{transASD}) more
transparent, we write the $(n-4)$ form $\Omega$ as
\beq
\Omega = \A (dz^1 \wedge \cdots \wedge dz^{n-4}) + \widetilde{\Omega}~.
\eeq
The first term is proportional to the volume form of the base space $\Sigma$ and
\beq
\widetilde{\Omega} |_\Sigma = 0~.
\eeq
By the definition of $*$-operator the
right hand side of (\ref{transASD})  has to be proportional to the volume form
of $\Sigma$. Hence, we must have
\beq
\A F_{yy} =  *_4 F_{yy}~,
\eeq
where $*_4$ stands for the Hodge dual on the transverse four dimensional
fibre. Since the square of the Hodge dual operator is one in four
dimensions, $\A =1$ and
we finally obtain
\beq
\Omega |_\Sigma = dz^1 \wedge \cdots \wedge dz^{n-4}~.
\eeq
Hence, $\Sigma$ has to be a SUSY $(n-4)$ cycle with respect to
the $(n-4)$ form $\Omega$. Moreover, we have
\beq
F_{yy} =  *_4 F_{yy}~,
\eeq
the four dimensional self-dual condition in the transverse direction !

We have made a partial four dimensional rescaling
to get an idea on the higher dimensional analogue 
of point like instantons. The picture we have obtained
is a family of four dimensinal instantons over a $(n-4)$
dimensinal SUSY cycle. Note that this is {\it not} 
a direct product of the SUSY cycle $\Sigma$ and self-dual instantons.
The moduli (e.g. the size and shape) of the instantons
may in principle depend on the coordinates
of the SUSY cycle $\Sigma$.  
To be mathematically more precise, the four dimensional
transverse direction should be identified with
the fibre in the normal bundle $N(\Sigma)$ of the SUSY
cycle $\Sigma$. Thus what we have seen is that higher
dimensional instantons can be topologically reduced
to those on the normal bundle $N(\Sigma)$ over 
an $(n-4)$ dimensional submanifold $\Sigma$,
which is necessarily an $\Omega$-calibrated
submanifold. Note that the dimensions of the total space of 
the normal bundle $N(\Sigma)$ is the same
as the original manifold $M^n$.
Our gauge theory is still defined in $n$-dimensions.
It is not to be confused with the dimensional
reduction to the submanifold $\Sigma$.

To illustrate the above idea let us look at
an explicit example. Some time ago Fubini-Nicolai and
Fairlie-Nuyts constructed an octonionic instanton 
\cite{FN1},\cite{FN2}. (See also \cite{IP}.)
Their solution is to be regarded as an instanton
on the flat eight dimensional Euclidean space ${\bf R}^8$
and its action is divergent as follows from a general
argument in section 2. However the octonionic instanton
of FNFN well illustrates some of the features
we have discussed. The octonionic instanton
equation is the generalized self-dual instanton equation 
(\ref{OASD}) in eight dimensions with the Cayley four form
\beq
\Omega = \frac{1}{4!} C_{abcd}~
e^a \wedge e^b \wedge e^c \wedge e^d~,
\eeq
where $C_{abcd}$ is defined in terms of 
the structure constants of the algebra of octonions
and $\{ e^a = e^a_\mu dx^\mu \}$ is an orthonormal frame (veilbein).
The four form $\Omega$ is self-dual ($* \Omega = \Omega$) in eight dimensions 
and closed on a $Spin(7)$ manifold \cite{Joy}. On such a manifold
there is a unique covariantly constant spinor $\epsilon$ and $\Omega$ is
also expressed by
\beq
 C_{\mu\nu\rho\sigma} = C_{abcd} e^a_\mu e^b_\nu e^c_\rho e^d_\sigma = 
\left( \overline{\epsilon}~\Gamma_{\mu\nu\rho\sigma}
\epsilon \right)~.
\eeq
The constant fourth rank tensor $C_{abcd}$ satisfies 
the following identity \cite{DGT} \cite{dWN};
\beq
C_{abst} C^{cdst} 
= 6 (\delta_a^c  \delta_b^d - \delta_a^d  \delta_b^c) 
-4 C_{ab}^{\ \ cd}~.
\eeq
Hence the linear map 
\beq
C : F \mapsto * (\Omega \wedge F)~,
\eeq
on the space of two-forms $\La^2({\bf R}^8)$ satisfies
\beq
(C - 2E)(C+6E) = 0~,
\eeq
and the eigenvalues of $(1/2) C$ are $1$ and $-3$.
We obtain the eigenspace decomposition;
\beq
\La^2({\bf R}^8) = E(1) \oplus E(-3)~.
\eeq
Since the linear map $C$ is traceless and $\hbox{dim}~\La^2 ({\bf R}^8)= 28$,
we have
\beq
\hbox{dim}~E(1) = 21, \quad \hbox{dim}~E(-3) = 7~.
\eeq
The octonionic instanton equation 
\beq
F_{\mu\nu}~=~\frac{1}{2} C_{\mu\nu\rho\sigma} F^{\rho\sigma}
\label{octinst}
\eeq
means the curvature has no components in $E(-3)$.

Let $G_{\mu\nu}$ be generators of $Spin(7)$. From an
ansatz for the gauge field;
\beq
A_{\mu} = G_{\mu\nu} \partial_\nu f~,
\eeq
we obtain the following solution to (\ref{octinst});
\beq
A_\mu = -\frac{2}{3}\frac{1}{\rho^2 + x^2} G_{\mu\nu} x_\nu~,
\eeq
where $x^\mu$ are Euclidean coordinates of ${\bf R}^8$ and $\rho$ is a parameter
in integration that represents the \lq\lq size\rq\rq\ of the
instanton. Using the commutation relation of $Spin(7)$ generators,
we can compute the curvature
\beq
F_{\mu\nu}
= \frac{2}{3}\frac{2 \rho^2 + x^2}{(\rho^2 + x^2)^2} G_{\mu\nu} 
 + \frac{4}{3} \frac{x_{[\mu} G_{\nu]\lm} x^{\lm}}{(\rho^2 + x^2)^2} 
- \frac{2}{9} \frac{x^\lm x^\sigma C_{\mu\nu\lm\kappa} G^\kappa_\sigma}
{(\rho^2 + x^2)^2}~. 
\eeq
Then the density of the second Chern class is
\beq
{\hbox {Tr}}~F_{[\mu\nu} F_{\lm\sigma]}
= \frac{3 \rho^2 + x^2}{(\rho^2 + x^2)^3} C_{\mu\nu\lm\sigma} 
+  \frac{4( 4 \rho^2 + x^2)}{(\rho^2 + x^2)^4} 
x_{[\mu} C_{\mu\nu\sigma] \kappa} x^\kappa~.
\eeq
Now let us choose a set of four indices $1 \leq M, N, P, Q \leq 8$, 
such that the component $C_{MNPQ}$ is non vanishing and 
consider a four dimensional plane $\Sigma$ extended to these directions.
Then $\Sigma$ is a flat non-compact SUSY cycle, or an $\Omega$-calibrated plane.
To follow the previous argument, we split the coordinates $x^\mu$ into
those of the plane $\Sigma$, $z^i~(i= M, N, P, Q)$ and the transverse
coordinates $y^\A$. Since ${\bf R}^8$ is topologically trivial,
the normal bundle $N(\Sigma)$ of the four dimensional plane $\Sigma$ is
again ${\bf R}^8$.  Due to the self-duality of $\Omega$ the coordinate
indices of
the base $\Sigma$ and the fibre are interchangeable.
With this special combination of four indices,
the density of the second Chern class is reduced to
\beq
{\hbox {Tr}}~F_{[MN} F_{PQ]}
= C_{MNPQ} 
\frac{3 \rho^4 + z^2( 4 \rho^2 + y^2 + z^2)}{(\rho^2 + y^2 + z^2)^4}~.
\label{density}
\eeq
Note that if we chose a four dimensional plane that is not calibrated,
then $C_{MNPQ}=0$ on the plane and we could not see 
such a topological density.
Finally the integration on the transverse coordinates gives
\beq
\int {\hbox {Tr}}~F_{[MN} F_{PQ]} d^4y
= \frac{\pi}{2} C_{MNPQ}~.
\eeq
We find that the second Chern class is independent of the scale
parameter $\rho$ and moreover of the coordinates $z^i$ of the SUSY cycle
$\Sigma$. Hence the topological type of the self-dual instantons
over the SUSY cycle is universal. This fact suggests that, if we could take
a compact base space $\Sigma$, there would be a chance for the total action
to be finite.
From (\ref{density}) we see that the \lq\lq shape\rq\rq
of the instantons does depend on the coordinates $z^i$ and the octonionic
instanton does not have a direct product structure.
But the dependence of the instanton moduli on the coordinates
of the SUSY cycle $\Sigma$ cannot be arbitrary.
From the result in \cite{FKS} on a topological
dimensional reduction of octonionic instantons,
we expect the triholomorphic condition would control such dependence.
In the limit of the scaling of the transverse coordinates $y^\A \rightarrow
\lm y^\A$
together with the scaling of the parameter $\rho \rightarrow \lm\rho$,
we recover the standard action density of one instanton solution;
\beq
{\hbox {Tr}}~F_{[MN} F_{PQ]} = \frac{3 \rho^4}{( \rho^2 + y^4 )^4}~.
\eeq
Thus after the rescaling of the four dimensional fibre, we find exactly
a four dimensional instanton.


\section{Conclusion and Outlook}

Higher dimensional instanton equation can be used in
constructing cohomological Yang-Mills theories and the geometry of
the moduli space may lead a new development
in mathematics. However, as was discussed in section 2,
there are a few issues in a physically acceptable interpretation
of these instantons as solutions in the higher dimensional
gauge theory. These are closely related to the fact that
the standard Yang-Mills action is conformal invariant 
only in four dimensions. Motivated by this fact, we have
paid attention to an $(n-4)$ dimensional submanifold $\Sigma$ and
made a scaling transformation along the four dimensional fibre
of the normal bundle $N(\Sigma)$. We have shown that $\Sigma$ 
has to be sypersymmetric and reduced higher dimensional
instantons to configurations on the total space of $N(\Sigma)$.
Though this picture of a fibration of four dimensional instantons
over a SUSY cycle gives only a local characterization of 
higher dimensional instantons, we believe that this is a good starting
point to discuss the moduli problem of higher dimensional
instantons. As we have emphasized before, this is
a natural analogue of point-like (ideal) instantons 
in four dimensions.

There is a nice formulation of small-size instantons  
in the frame-work of D-branes and this also explains
the ADHM construction of four dimensional instantons \cite{Wit}.
It is an interesting challenge to develop a similar formulation in terms of
D-brane
for a family of instantons over a SUSY cycle. Eventually 
a resolution of the issues mentioned above might be obtained 
by embedding the instantons in the brane picture of string theory or $M$ theory.

In algebraic geometry a compactification of 
the instanton moduli space can be discussed as
the problem of holomorphic vector bundles.
It is interesting to compare
the construction discussed in this paper, which is differential
geometrical, with the idea from algebraic geometry.
The case of holomorphic vector
bundles on Calabi-Yau three-folds or four-folds seems to
be a good example for the comparison.

\vskip10mm

\begin{center}
{\bf Acknowledgements}
\end{center}

I am grateful to the organizers of the workshop for providing
a chance to give a talk. Among the participants I would like
to thank E. D'Hoker, T. Eguchi, M. Mari\~no and Y. Yasui for discussions
and comments. The initial stage of this work was done, when I visited
MIT last summer. I would like to thank G. Tian for illuminating
discussion and I.M. Singer for warm hospitality.

This work is supported in part by the Grant-in-Aid
for Scientific Research on Priority Area 707 
"Supersymmetry and Unified Theory
of Elementary Particles" and No. 10640081,
from Japan Ministry of Education.

\newpage


\end{document}